\documentclass[traditabstract]{aa}
\usepackage{epsfig,graphicx}
\usepackage{natbib}
\usepackage{color}
\bibpunct{(}{)}{;}{a}{}{,}    

\begin{document}

\title{Calibration of the 6302/6301 Stokes $V$ line ratio\\ in terms of the 5250/5247 ratio}
\titlerunning{Calibration of the  6302/6301  Stokes $V$ line ratio}
\authorrunning{J.O. Stenflo et al.}
\author{J.O. Stenflo$^{1,3}$\and  M.L. Demidov$^{2}$\and M. Bianda$^{1}$\and R. Ramelli$^{1}$}

\institute{Istituto Ricerche Solari Locarno (IRSOL), Via Patocchi, CH-6605 Locarno, Switzerland
 \and Institute of Solar-Terrestrial Physics, Siberian Branch, Russian Academy of Sciences, P.O. Box 291,\\  664033 Irkutsk,  Russia \and 
Institute of Astronomy, ETH Zurich, CH-8093 Zurich, Switzerland}

\date{}

\abstract{ 
Four decades ago the Stokes $V$ line ratio in the Fe\,{\sc i} 5247.06 and 5250.22\,\AA\ lines was introduced as a powerful means of exploring the intrinsic field strengths at sub-pixel scales, which led to the discovery that most of the photospheric flux is in intermittent kG form. The ``green'' 5247-5250 line pair is unique because it allows the magnetic-field  effects to be isolated from the thermodynamic effects. No other line pair with this property has since been identified. In recent years much of the magnetic-field diagnostics has been based on the ``red'' Fe\,{\sc i} 6301.5 and 6302.5\,\AA\ line pair, since it was chosen in the design of the \textit{Hinode} space observatory.  Although thermodynamic effects severely contaminate the magnetic-field signatures for this line ratio, it is still possible to use it to extract information on intrinsic magnetic fields, but only after it has been ``renormalized'', since otherwise it produces fictitious, superstrong fields everywhere. In the present work we explore the joint behavior of these two line ratios to determine how the ``contaminated'' red line ratio can be translated into the corresponding green line ratio, which then allows for a direct interpretation in terms of intrinsic magnetic fields. Our observations are mainly based on recordings with the ZIMPOL-3 spectro-polarimeter at IRSOL in Locarno, Switzerland, complemented by data from the STOP telescope at the Sayan solar observatory (Irkutsk, Russia). The IRSOL observations are unique by allowing both the green and red line pairs to be recorded simultaneously on the same CCD sensor. We show how the line ratios depend on both the measured flux densities and on the heliocentric distance (the $\mu$ value on the solar disk), and finally derive the calibration function that enables the red line ratio to be translated to the green ratio for each $\mu$ value. 
}
\maketitle

\section{Introduction}\label{sec:intro}
The techniques used for the determination of active-region magnetic fields, like the various forms of Stokes inversions, cannot in general be used to diagnose quiet-sun magnetic fields. The two main reasons for this are the low S/N ratio of the measured polarization signatures on the quiet Sun, combined with the circumstance that the field is structured on scales far smaller than the telescope resolution. Reliable magnetic-field information can only be obtained with specially designed methods that are robust with respect to measurement noise and insensitive to model assumptions. 

The application of such robust techniques is of central importance to solar physics, since the ``quiet Sun'' represents most of the solar surface, almost all of it during periods of low solar activity. Quiet solar regions are generally the main sources of open solar magnetic flux and, as a consequence, of the interplanetary magnetic field. 
  
While many different spectral lines have been used for magnetic-field recordings, it is not possible to obtain information on the intrinsic field strengths of quiet-sun magnetic fields from observations in single spectral lines. Such information can only be derived from simultaneous observations in carefully chosen combinations of spectral lines \citep[cf.][]{DemBalt12, BaltDem12}.

Successful diagnostics depends on an optimum choice of line combinations. During recent years, in particular because of the great success of the \textit{Hinode} space mission \citep{Suematsu-etal08}, special attention has been given to the diagnostic possibilities with the red line pair Fe\,{\sc i}~6301.5 and 6302.5\,\AA\ \citep{
MartinezGonzalez-etal06, stenflo-litesetal08,Stenflo10, Stenflo11, Stenflo12}. 
By applying the line-ratio method to the  \textit{Hinode}  SOT/SP data two distinct magnetic flux populations were discovered \citep{Stenflo10, Stenflo11}, corresponding to strong (collapsed) and weak (uncollapsed) flux. However, since this line pair is not well suited for the diagnostics of intrinsic magnetic field strengths, a special ``renormalization procedure'' had to be devised \citep{Stenflo10} before the line ratio could be interpreted. Although  these two lines belong to the same multiplet, they  differ not only in the values of their Land\'e factors ($g$ = 1.667 for Fe\,{\sc i}~6301.5\,\AA, $g$ = 2.5 for Fe\,{\sc i}~6302.5\,\AA), but also in their line strengths and line-formation properties. Therefore the magnetic-field information in their Stokes $V$ line ratio is contaminated by thermodynamic effects, as the two lines are formed in different atmospheric layers \citep[cf.][]{Faurobert-etal10}. 

To eliminate such contamination in the data analysis one needs to relate the red 6302.5/6301.5 line ratio to the green 5250.22/5247.06 line ratio \citep{Stenflo73}, which does not suffer from this problem, since  the spectral lines Fe\,{\sc i}~5247.06\,\AA\ and Fe\,{\sc i}~5250.22\,\AA\  not only belong to the same multiplet but in particular also 
have the same line strengths and excitation potentials. They are therefore formed in the same way and in the same locations of the solar atmosphere, the only difference being their Land\'e factors. If the spatially unresolved fields were intrinsically weak ($\la 500$\,G), the ratio between their Stokes $V$ amplitudes would scale in proportion to their Land\'e\ factors. For stronger intrinsic fields, however, nonlinear effects due to Zeeman saturation will cause the $V$ ratio to differ from the weak-field ratio. Such differential nonlinear effects have nothing to do with how much flux there is in the measured region, they are exclusively a function of the field strength, regardless of the sizes of the subresolution elements. They are therefore independent of telescope resolution. The measured green line ratio can only differ from its weak-field value if the ``hidden'' (subpixel) magnetic structures are truly strong (of order kG), in contrast to the red line ratio, which can differ from the expected weak-field ratio because of contamination from differential thermodynamic effects. The first application of the green line ratio led to the discovery that more than 90\% of the quiet-sun magnetic flux seen in magnetograms with moderately high spatial resolution (a few arcsec) has its origin in strong, intermittent kG flux bundles or flux tubes, although the apparent magnetic flux densities in the magnetograms are typically two orders of magnitude smaller \citep{Stenflo73}. 

According to \citet{Socas-Navarro-etal08} the 6302.5/6301.5 line ratio is as useful as the 5250/5247 line ratio for Stokes inversion of observations with S/N ratios sufficiently high for the Stokes inversion approach to be applicable (which is rarely the case in quiet regions). On the other hand \citet{{stenflo-khocoll07}} used MHD simulations to compare the performance of these two line ratios and infrared lines with large Zeeman splitting. They  concluded that the red 6302.5/6301.5 line ratio, if interpreted directly, systematically produces fictitious, superstrong fields, while the field is correctly diagnosed by the 5250/5247 line ratio and by the infrared lines. They further identified the origin of the fictitious results linked to the red line ratio as being due to spatially unresolved correlations between magnetic and velocity field gradients along the line of sight (as in particular occurs across canopies). The ``renormalization'' of the red line ratio \citep{Stenflo10} was introduced as a way to circumvent this problem so that the line-ratio technique could be applied to the \textit{Hinode} observations. In the present paper we explore the validity of this renormalization procedure. 

Since the outstanding \textit{Hinode} space telescope unfortunately was not designed to include the ``ideal'' 5247-5250 line pair, there is a need to explore how the ``non-ideal'' 6301.5-6302.5 line pair used not only by \textit{Hinode} but also by SOLIS/SVM  \citep{Henney-etal09} behaves and relates to the 5247-5250 line pair, and to cross-calibrate the two line pairs. We approach this objective by making observations, separate and simultaneous, in both combinations of spectral lines, the ``green'' and the ``red'' ones, to explore the magnitude of the thermodynamic contamination of the   6302.5/6301.5 line ratio. A special observing program for this was carried out at the IRSOL (Istituto Ricerche Solari Locarno) Gregory Coud\'e Telescope with  ZIMPOL-3 (Zurich IMaging POLarimeter) \citep{Ramelli-etal10} during the summer of 2012. ZIMPOL is currently the worldwide most sensitive polarimeter for solar observations in the visible part of the spectrum. The last-generation version at IRSOL allows unprecedented polarimetric quality. A major advantage of the IRSOL telescope system is that its spectrograph uses exclusively reflection optics. It is therefore free from chromatic aberrations, which allows both the green and red line pairs, which can be made to overlap in two particular grating orders, to be recorded simultaneously, side by side, on the same CCD sensor of the ZIMPOL-3 camera system. 

\begin{figure*}
\centering
\resizebox{0.8\hsize}{!}{\includegraphics{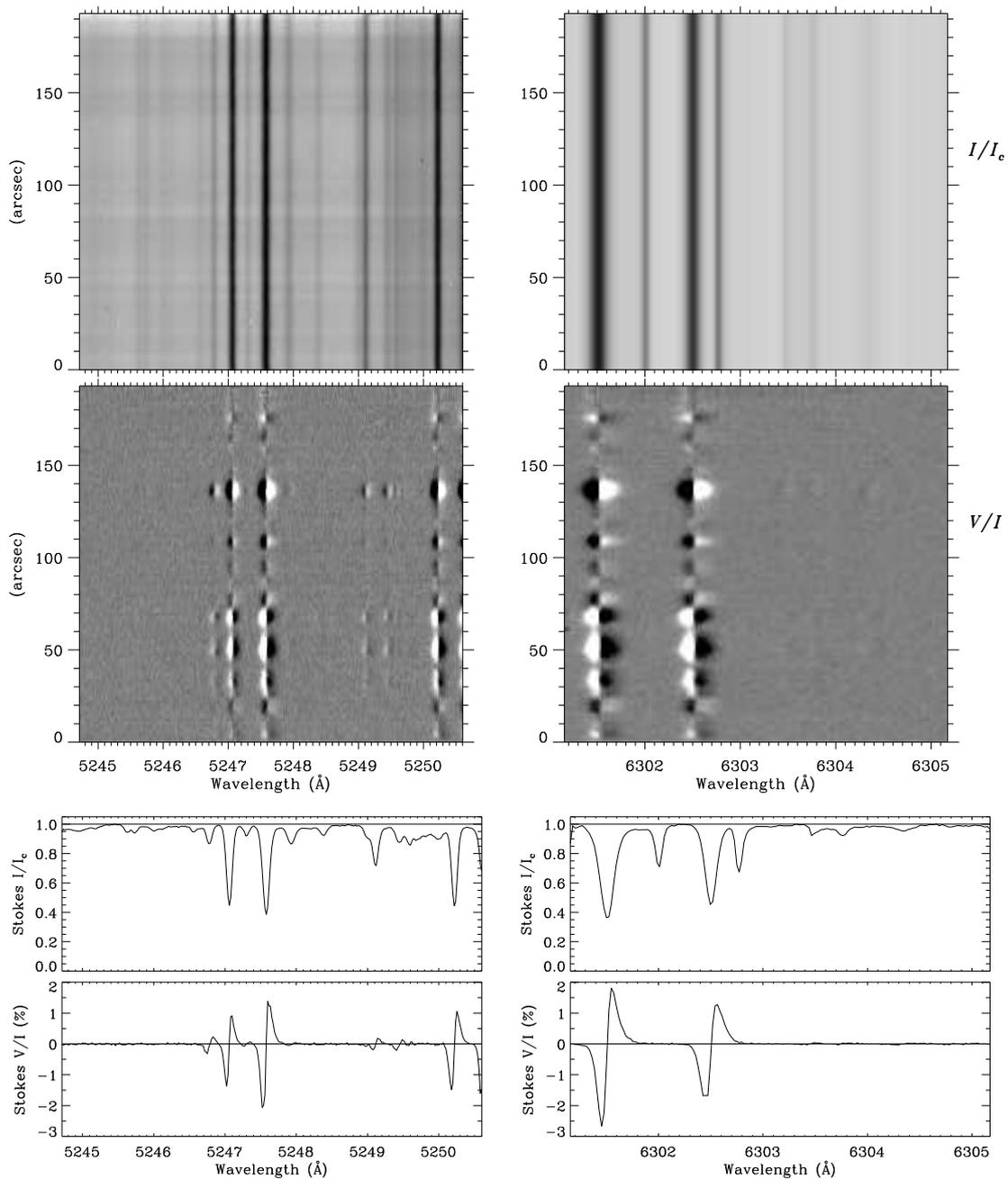}}
\caption{Example of Stokes spectra of the ``green'' Fe\,{\sc i}~5247.0 and 5250.2\,\AA\ (left panels) and ``red'' Fe\,{\sc i}~6301.5 and 6302.5\,\AA\ (right panels) lines recorded simultaneously on the same CCD sensor. The first two rows of panels show the 2-D images of Stokes $I/I_{c}$ and $V/I$. The grey-scale cuts for $V/I$ have been set low, at $\pm 0.3$\,\%, to better bring out the weak flux densities. The last two rows of panels show $I/I_{c}$  and $V/I$ for the CCD row that is located at  135\,arcsec. The recording was made with ZIMPOL-3 at IRSOL at the center of the solar disk on August 11, 2012. 
}\label{F1-ccd-2filt-together}
\end{figure*}

An example of such simultaneous observations is shown in Figure \ref{F1-ccd-2filt-together} for one typical recording. The top panels show the 2-D spectral images of $I/I_{c}$, the second row of panels the Stokes $V/I$ images of the fractional circular polarization. One can see that $V/I$ is highly structured along the slit, with no indication of a diffuse background field. The maximum $V/I$ signals in this particular recording are found in CCD row number 95 (at $\approx 135^{\prime\prime}$ along the spatial scale). The Stokes $I/I_{c}$ and $V/I$ spectra for this CCD row are shown in the third and fourth rows of panels.  
 
A significant fraction of our IRSOL observations were made in such ``simultaneous'' mode, which allowed  us to  study the pixel-to-pixel correlations between the polarimetric signals in the red and the green lines. However, because the choice of grating orders imposed by the requirement of simultaneous observations is not optimum and requires a more complicated arrangement in the focal plane, we also made recordings in ``non-simultaneous'' mode, separately for the red and green lines.  
 
Such (separate for the two wavelength ranges) type of observations is the only possible observing mode for the STOP (Solar Telescope for Operative Predictions) telescope at the Sayan Solar Observatory (SSO), because the use of lens optics in the Littrow spectrograph \citep{Demidov-etal02} causes the location of the spectral focus to be significantly different in the red and green ranges. Nevertheless the STOP  full-disk magnetograms in the ``red'' and ``green'' lines were also analyzed  to obtain independent and complementary information on the Stokes $V$ line ratios in different combinations of spectral lines.

\section{Observations at solar disk center with ZIMPOL-3}\label{S-ObsCenter}     

\subsection{Instrumental set-up}\label{sec:instr}   
Since we used a special set-up at IRSOL for the present project, it needs to be described in some detail here. 
 
 As mentioned above, two observing modes were used, ``simultaneous'' and ``separate''. In the case of the separate mode, the order-separating interference filter is placed behind the spectrograph slit. The 11th  order is then used for the green lines, the 9th order for the red lines. In the green range the  spectral resolution is 7.8\,m\AA\ pro pixel of the ZIMPOL-3 camera, in the red range it is  6.8\,m\AA\ pro pixel. The effective integration time used (to accumulate 600 images) was  about 10 minutes for both spectral ranges.

The wavelength ratio of the two line pairs is very close to the integer ratio $5/6$. The two ranges therefore overlap in the spectrograph focal plane if the red lines are observed in the 5th order while the green lines are observed in the 6th order. To separate the overlapping orders, two square-shaped interference filters are placed side by side in the spectrograph focal plane,  without any interference filter at the entrance slit. A rectangular area of the spectrograph focal plane is re-imaged by an achromatic telecentric image reducer 
on the CCD sensor of the ZIMPOL-3 camera, allowing simultaneous recording of the two 
line pairs. In this mode the spectral resolution is 19.3\,m\AA\ pro pixel for the green line pair, 23.2\,m\AA\  pro pixel for the red pair. The effective integration time (required to accumulate 400 images) was about 4 min (since a smaller sub-frame was used as compared with the full frame recording for the separate mode).

For both observing modes the spectra have the same spatial resolution corresponding to 1.44\,arcsec  pro pixel, which corresponds to a resolution comparable to the typical seeing smearing over the extended exposure times used.

The spectrograph slit used was 0.06\,mm wide, which corresponds to about 
0.5\,arcsec in the solar image. It was chosen that narrow in order not to compromise the 
spectral resolution.

\begin{figure*}
\centering
\resizebox{0.9\hsize}{!}{
\includegraphics{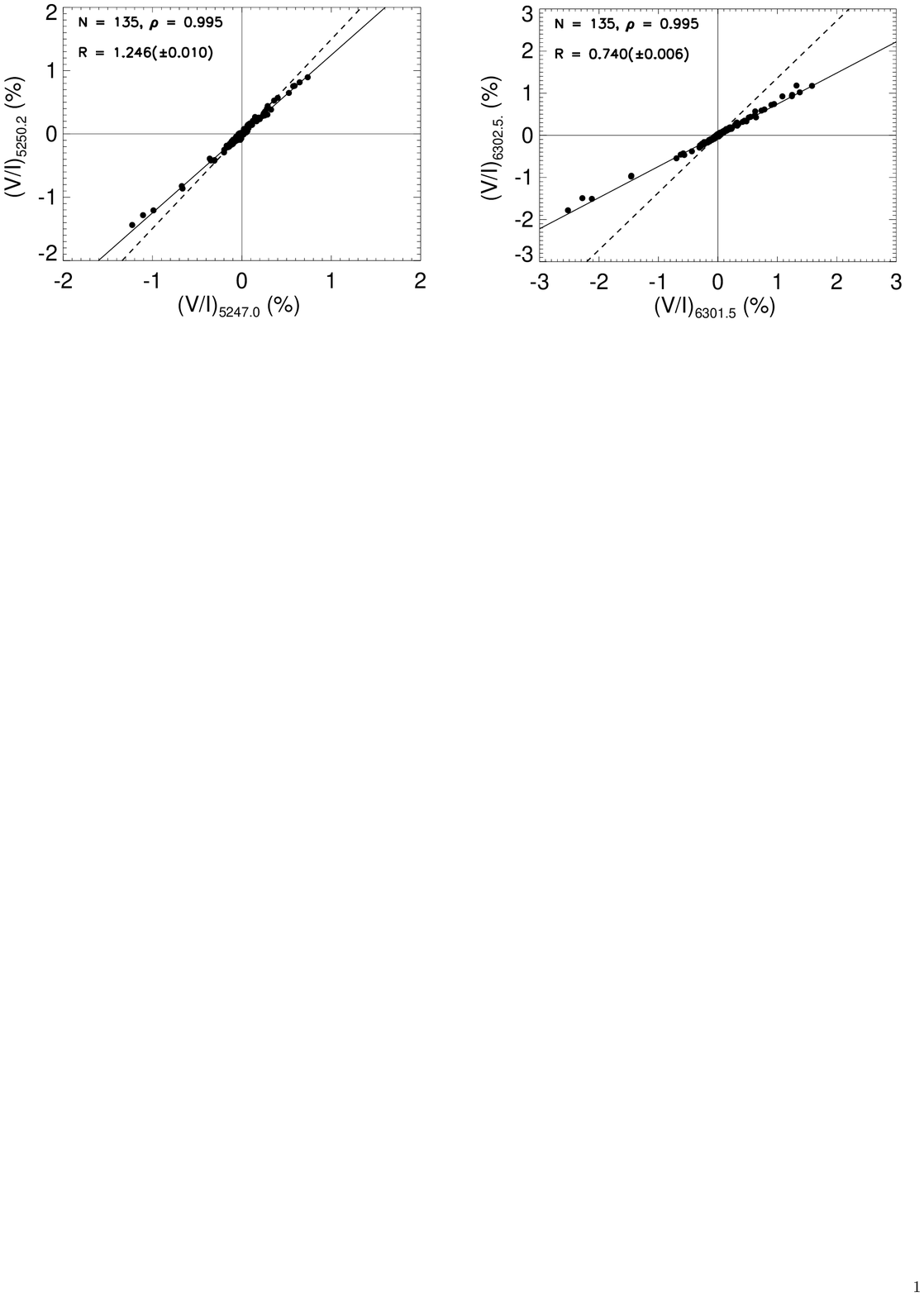}}
\caption{$V/I$ scatter plots for the green (left panel) and red (right panel) line pairs for the recording that was shown in Figure \ref{F1-ccd-2filt-together}, which represented a single  recording in ``simultaneous'' mode. The solid lines are linear regression fits, while the dashed lines represent the weak field slopes (WFS). $N$ is the number of spatial pixels that are represented, $\rho$ is the correlation coefficient, $R$ the regression-line slope. 
}\label{F2-g2g1-r2r1-2filt}
\end{figure*}

In a Gregory Coud\' e telescope the solar image, which is projected on the focal 
plane with the entrance slit of the spectrograph, rotates in synchrony 
with the rotation of Earth. To prevent image rotation during an extended integration and to allow the alignment of the slit on the solar image in any desired direction, an 
image derotator (a Dove prism) can be placed in front of the slit. Since the  
derotator is  located after the polarization analyzer (modulator 
plus linear polarizer) the polarization 
measurement is not affected by the polarizing properties of the rotator. The Dove prism is however chromatic: it can cause a slight relative shift of the 
two monochromatic images. For some configurations the relative spatial shift between the  red and green spectra corresponds to about three CCD 
pixels on the ZIMPOL-3 camera. For this reason we decided not to use the 
derotator for the simultaneous observing mode, to
avoid such chromatic effects. A disadvantage of this is the absence of 
image rotation compensation. This means that at the extreme ends of the slit in the spectral image during the effective 240\,s integration time there is a rotational smearing of slightly more than 1\,arcsec, but as this remains below the typical effective smearing by atmospheric seeing, it is acceptable. 

For the polarization modulation we used a photoelastic modulator (PEM). Its oscillation 
amplitude needs to be set such that the modulation efficiency at a selected wavelength gets 
optimized.  To simultaneously optimize the efficiency in the green and red, the modulation efficiency was maximized for 5800\,\AA\ as a compromise setting. The consequence of not setting the PEM for exactly the measured wavelength only implies a small loss in efficiency 
but otherwise not a loss in quality (it just requires a longer integration time to 
reach a given polarimetric noise level). 

When doing the center-to-limb observing runs, the correct position on the solar disk was reached with the help of the primary image guiding system of the telescope \citep{Kueveler11}. Close to the limb, for $\mu=\cos\theta\la 0.35$, we in addition used a computer-controlled tilt-plate to stabilize the chosen distance of the spectrograph slit from the solar limb. The data reduction pipeline was basically the same for the simultaneous and separate modes of observation, and included calibration, subtraction  of dark current, and flat field correction.  

\subsection{Extraction of the $V/I$ amplitudes}\label{sec:extract}
Note that in all ZIMPOL observations we work with the fractional polarizations, thus in the case of the circular polarization with $V/I$, not with $V/I_c$. The reason is that each of Stokes $V$ and Stokes $I$ is affected by relatively large pixel-to-pixel gain-table variations, vignetting and other optical imperfections, but since all these effects are identical for $V$ and $I$, they divide out completely when forming the fractional polarization. This is not at all the case with $V/I_c$. The polarimetric precision that is possible with ZIMPOL ($10^{-5}$ if the number of photoelectrons required by Poisson statistics are collected) can only be reached in terms of the fractional polarization. 

Due to the low S/N ratios of polarimetric recordings on the quiet Sun with a large fraction of the spatial pixels having $V/I$ amplitudes of order $10^{-4}$, comparable to the measurement noise, it is imperative to use a technique for extraction of the observed amplitudes that is both robust and gives an unbiased, symmetric Gaussian noise distribution, regardless of how small the S/N ratios are. Such a technique was developed and successfully implemented for the analysis of the Hinode SOT/SP 6301.5 - 6302.5 Stokes spectra by \citet{Stenflo10,Stenflo11}, and it is the technique we will use here as well. 
 
\begin{figure*}
\centering
\resizebox{0.9\hsize}{!}{
\includegraphics{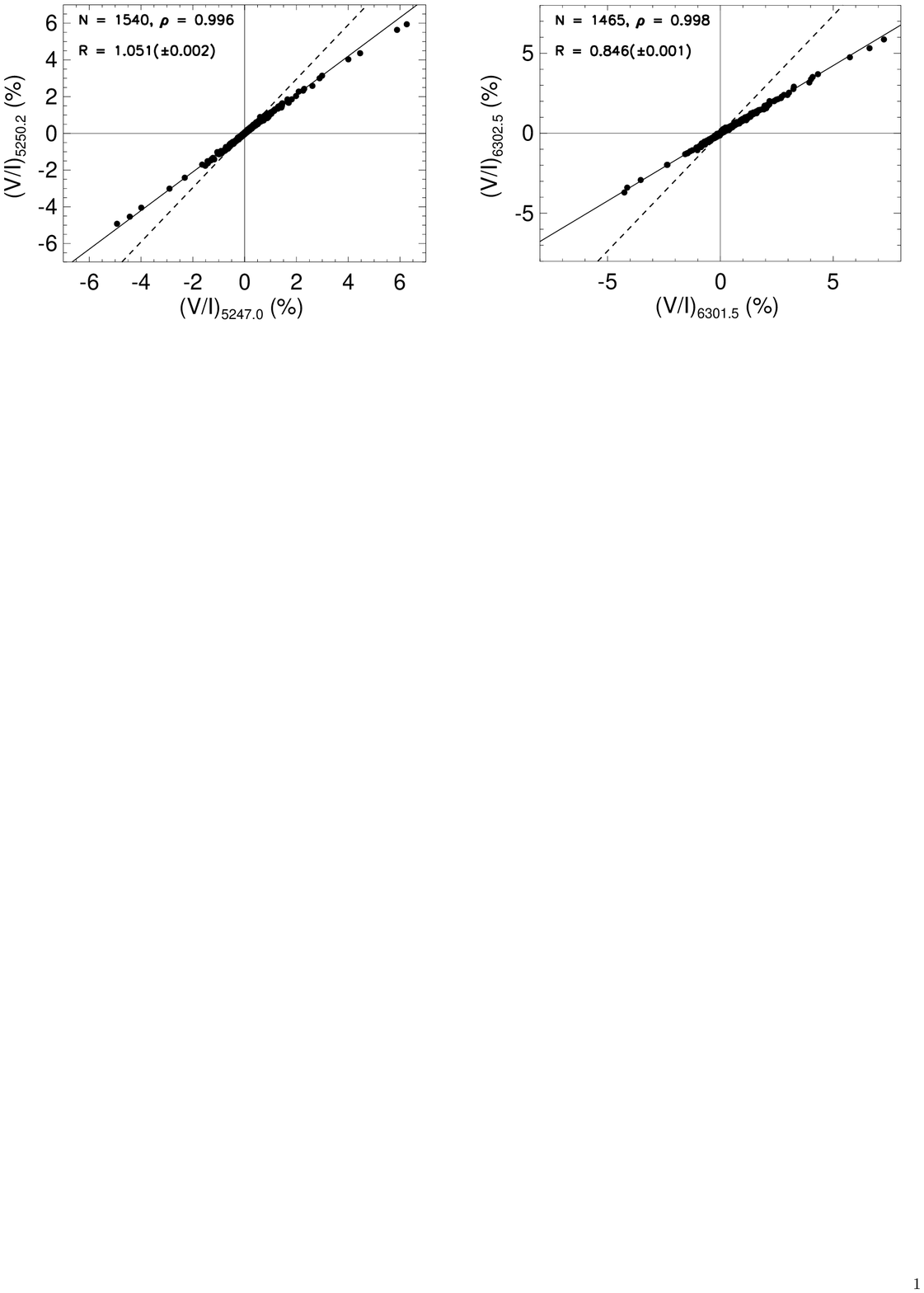}}
\caption{$V/I$ scatter plots like in Figure \ref{F2-g2g1-r2r1-2filt} but for the statistical sample represented by 11 disk-center recordings in ``separate'' mode.
}\label{F3-diskcentr-scat}
\end{figure*}

With this method the $V/I$ spectra for each spatial pixel is fitted to templates. To construct the templates a spatially averaged Stokes $V/I$ spectrum with high S/N ratio (because of the spatial averaging) is derived. From this average spectrum the blue and red lobes in the wings of the $V/I$ profiles of the respective spectral line are cut out and normalized, to be used as individual templates for the fitting process. After the Stokes spectrum from a given spatial pixel has been wavelength shifted (to compensate for its individual Doppler shift) and interpolated to the wavelength scale of the templates, the $V/I$ spectrum is fitted to the various lobe templates with an iterative least squares technique, with the lobe amplitude as the single free parameter. The fitting is extremely robust with immediate and unique convergence, regardless of how noisy the spectrum is, and it delivers the 1-$\sigma$ error estimate of the determined amplitude. Since the relation between the observed $V/I$ and the parameter to be fitted is linear, the error distribution is Gaussian. Although both the blue and red $V/I$ lobes are fitted separately, we will \citep[like in][]{Stenflo10} only make use of the blue lobe amplitudes for our line-ratio analysis, for the reason that the blue lobes on average have a shape agreeing with that of $\partial I/\partial\lambda$ (in contrast to the red lobes), indicating that they are less influenced by the distortions from subresolution correlations between the magentic and velocity gradients along the line of sight. In addition the blue lobes have on average a somewhat larger amplitude than the red lobes as a result of the Stokes $V$ asymmetries induced by the mentioned magnetic-velocity field correlations.

\subsection{Scatter plots}\label{sec:scatter}
 Scatter plots  of $V/I$ for the green and red line pairs for the recording that was illustrated in Figure \ref{F1-ccd-2filt-together} are shown in Figure \ref{F2-g2g1-r2r1-2filt}. The solid straight lines represent linear regressions calculated with the major axis method \citep{Davis-1986}.  The dashed lines give the slopes expected in the weak field case (in the following referred to as WFS, the ``weak field slope''). WFS represents the  ratio between the intensity gradients of the corresponding spectral lines, scaled in proportion to their Land\'e factors.  We expect the ratios of the Stokes $V/I$ signals to equal WFS in the case that the intrinsic magnetic field strengths are sufficiently small ($\la 500$\,G) for Zeeman saturation effects to be negligible.    
   
To collect more data to enhance the statistical analysis,  usually eleven recordings were made, with the spectrograph slit shifted (perpendicular to the slit direction) by about 20\,arcsec between  each recording. A similar number of recordings at spatially independent positions were carried out not only for the case of the disk center observations ($\mu = \cos\theta = 1.0$, where $\theta$ is the heliocentric angle), discussed in the present section, but also for other center-to-limb positions (different $\mu$ values). The center-to-limb variations will be discussed in Sect.~\ref{S-CLV LR}. 

Figure \ref{F3-diskcentr-scat} shows an example of scatter plots representing a collection of 11 recordings in separate mode at disk center. Inspection of both Figs.~\ref{F2-g2g1-r2r1-2filt} and \ref{F3-diskcentr-scat} reveals that all the regression slopes are significantly smaller than the WFS, but that the deviation from WFS is substantially larger for the red line pair as compared with the green pair. However, this does not imply that the hidden magnetic fields seen by the red lines are stronger. While in the case of the green line pair the deviation from WFS must originate from Zeeman saturation alone, in the case of the red line pair there is an additional unconstrained contribution from differential thermodynamic effects, as was demonstrated with the much higher spatial resolution of \textit{Hinode} \citep{Stenflo10}. 

Besides this difference between the green and red line pairs, which is brought out by both Figs.~\ref{F2-g2g1-r2r1-2filt} and \ref{F3-diskcentr-scat}, there is also a small slope difference between the two figures. Part of this difference has to do with the circumstance that the range of $V/I$ values covered by the different statistical samples is different, and that the assumption of a linear regression relation is not entirely valid, but is only an approximation. As we will see in Sect.~\ref{sec:fluxdep} below, the line ratio is not independent of $V/I$ amplitude, as it should be with the linear assumption, but the deviation from the WFS increases with increasing $V/I$. 

As the intrinsic regression relation is nonlinear with a $V/I$ dependence, our present approach of calculating regression slopes that is based on a linear assumption will lead to scatter in the derived slope values due to the varying ways in which the $V/I$ values get weighted. The reason why we have anyways chosen to use the linear assumption is that it is statistically very robust, and the deviations from linearity lead to effects that are relatively small in comparison with the main effect (slope deviation from WFS) that we are dealing with. In future work one may consider to use fully nonlinear fitting functions, but this would require significantly improved statistics and lies beyond the scope of the present exploratory work. 

\begin{figure*}
\centering
\resizebox{0.9\hsize}{!}{
\includegraphics{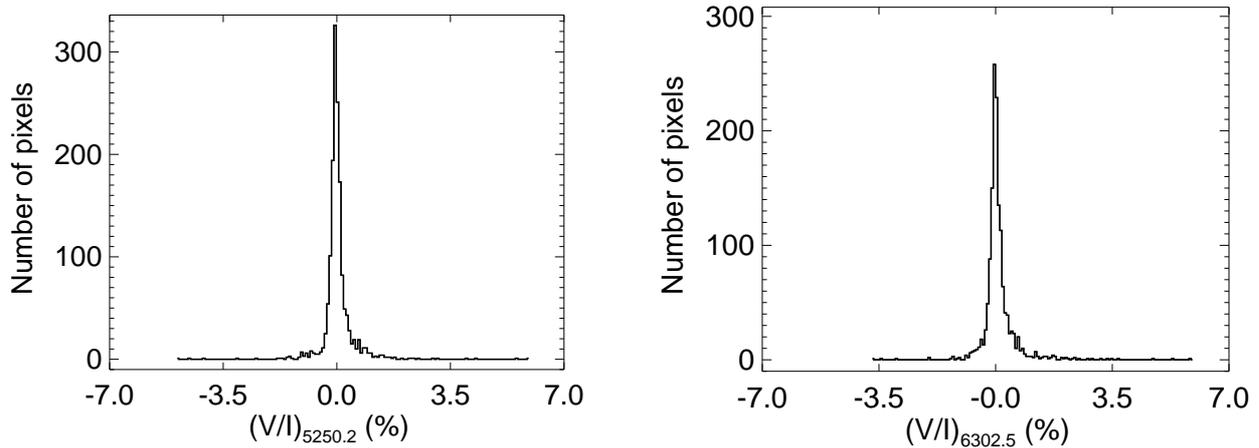}}
\caption{$V/I$ histograms for Fe\,{\sc i}~5250.2\,\AA\  (left panel) and Fe\,{\sc i}~6302.5\,\AA\ (right panel), for the sample of 11 disk-center recordings (from June 27, 2012) with the scatter plots that were illustrated in Figure \ref{F3-diskcentr-scat}.
}\label{F4-V-histograms}
\end{figure*}

\subsection{Gaussian fitting of line-ratio histograms}\label{sec:gauss}
Since all our observations represent quiet solar regions, and we have used the slit-jaw H$\alpha$ CCD camera to verify the absence of any obvious magnetic activity, all our statistical samples represent regions with very weak polarization signals. The $V/I$ histogram distributions (probability density functions) all show the characteristic sharp peak centered at zero polarization (representing zero flux density), with damping wings, which are suppressed and do not extend very far because all sampled regions were so quiet. Figure \ref{F4-V-histograms} shows for the 5250.2 and 6302.5\,\AA\ lines the histograms for the sample of 11 disk-center regions recorded in separate mode, whose scatter plots were shown in Figure \ref{F3-diskcentr-scat}. We see that the great majority of pixels have typical polarizations of order 0.1\,\%. 

\begin{figure*}
\centering
\resizebox{0.9\hsize}{!}{
\includegraphics{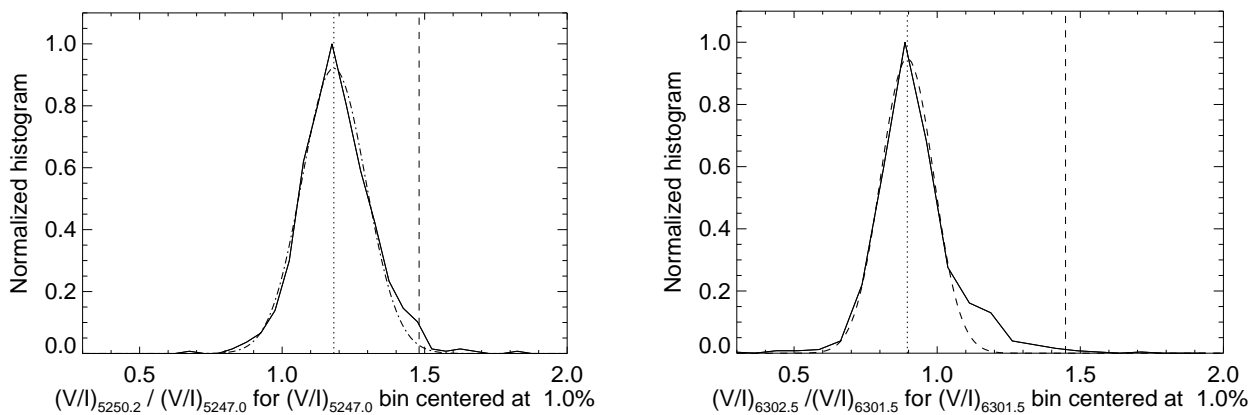}}
\caption{$V/I$ line-ratio histograms (solid lines) with Gaussian fit functions (dashed curves) for the same observational sample that was used in Figs.~\ref{F3-diskcentr-scat} and \ref{F4-V-histograms}, for $\vert V/I\vert$ bins  of $1.0\pm 0.9$\,\%. The positions of the Gaussians, marked by the vertical dotted lines, are much below the WFS positions, marked by the vertical dashed lines. Left panel: green line pair, representing a total of 731 data points. Right panel: red line pair, representing a total of 822 data points. 
}\label{F5-hist-gauss-fit}
\end{figure*}

\begin{figure*}
\centering
\resizebox{0.9\hsize}{!}{
\includegraphics{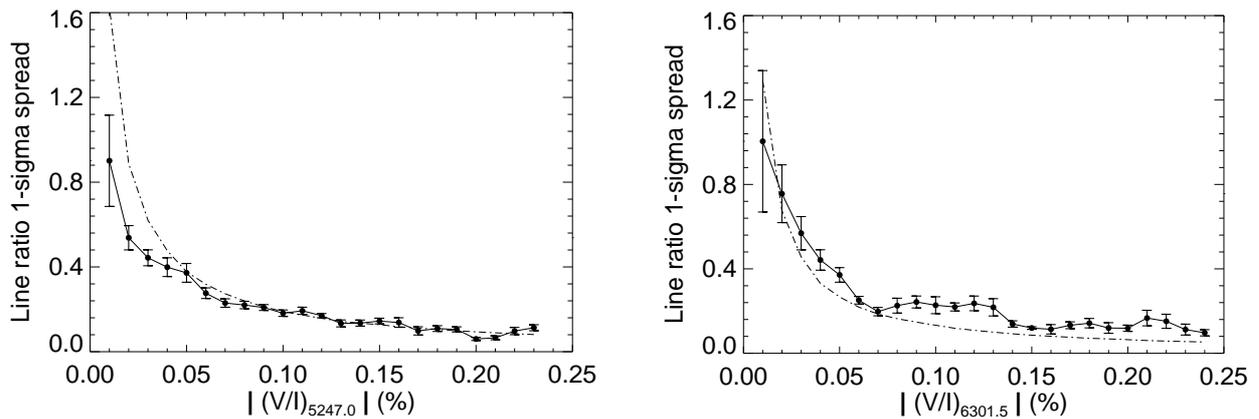}}
\caption{Variation of the standard deviation of the Gaussian fit function with $V/I$  amplitude, for the same observational sample that was used in Figs.~\ref{F3-diskcentr-scat} and \ref{F4-V-histograms} (solid lines). The dot-dashed curves represent the standard deviations that would result from the known measurement noise alone. The agreement with the solid curves is evidence that practically all the spread of the points around the regressions in the scatter plots is instrumental, with no discernible solar contribution. The left panel represents the green line pair, the right panel the red line pair. 
}\label{F6-1sigma-absV} 
\end{figure*}

\begin{figure*}
\centering
\resizebox{0.9\hsize}{!}{\includegraphics{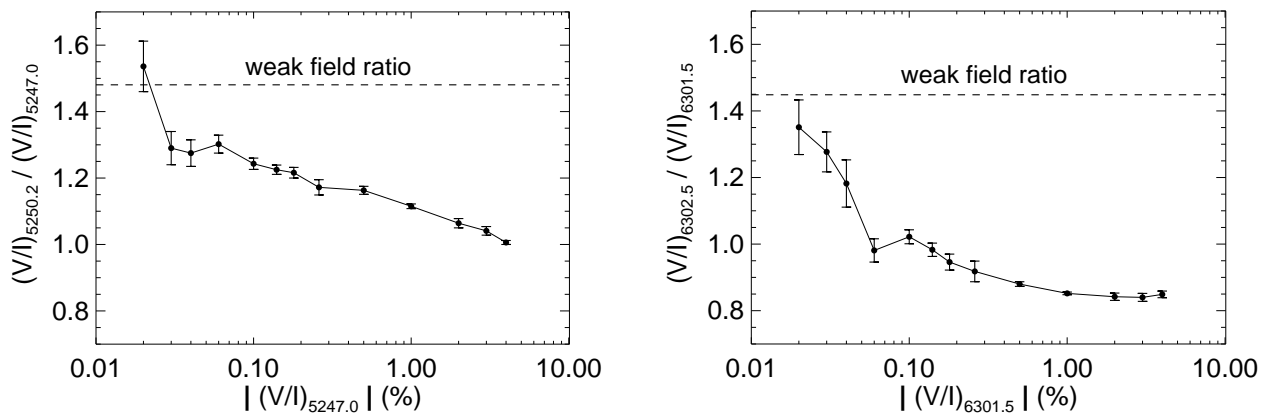}}
\caption{Variation of the $V/I$ line ratio with $V/I$ amplitude, as determined for each separate $\vert V/I \vert$ bin, for the same observational sample that was used in Figs.~\ref{F3-diskcentr-scat} and \ref{F4-V-histograms}. The left panel represents the green line pair, the right panel the red line pair. Due to the steep increase of the curve towards the WFS level in the limit of small $\vert V/I \vert$, a logarithmic scale is used for the horizontal axis.
}\label{F7-rat-absV-wide}
\end{figure*}

Besides the flux density (or $V/I$) histograms we can also produce line-ratio histograms for given bins in $\vert V/I\vert$ and do Gaussian fitting of the histogram functions for each such bin. The free fit parameters are the amplitude, position, and 1-$\sigma$ spread (standard deviation) of the Gaussian function. Figure \ref{F5-hist-gauss-fit} illustrates this Gauss fitting for the same observational sample that was used for Figs.~\ref{F3-diskcentr-scat} and \ref{F4-V-histograms}. The observed histograms are represented by the solid lines while the fitted Gaussian functions are dashed. The dotted vertical lines mark the position of the Gaussian, the dashed vertical lines the WFS location. The left panel for the green line pair is for a $\vert V/I\vert$ 5247 bin with an upper and lower boundary at $1.0\pm 0.9$\,\%, the right panel for the red line pair for a $\vert V/I\vert$ 6301.5 bin of $1.0\pm 0.9$\,\%. The bin width has for this example been chosen particularly wide to enhance the statistics and allow a detailed comparison between the shapes of the histograms and the Gaussian fit functions, in particular to bring out a secondary flux population seen as a bump in the distribution (see below). 

As a major fraction of the pixels have $V/I < 0.1$\,\%, one may wonder why we have not chosen to illustrate histograms in this range instead. The reason is that for very small $V/I$ the histograms spread and develop a complex appearance much less distinct from the WFS, because the standard deviation escalates with decreasing $V/I$ (cf. Fig.~\ref{F6-1sigma-absV} below) at the same time as the secondary flux population becomes prominent and merges with the main population. These complications are avoided with the choice of bins with larger $V/I$. 

The total number of data points that make up the histogram in the left panel of Fig.~\ref{F5-hist-gauss-fit} is 731, while for the right panel this number is 822. Since the total number of data points in the sample used are 1540 and 1465, respectively, according to Fig.~\ref{F3-diskcentr-scat}, approximately half of the surface area is represented by the histograms in Fig.~\ref{F5-hist-gauss-fit}.

Figure \ref{F5-hist-gauss-fit} shows again that the observed line ratio is much smaller than the WFS value, and that the deviation from WFS is substantially larger in the case of the red line pair. While the WFS deviation represents evidence for intrinsic kG fields as the source of most of the magnetic flux in the case of the green line pair, the red line ratio is contaminated by thermodynamic effects, which prevents a direct interpretation in terms of magnetic field strengths. 

For the green line pair in the left panel we notice evidence for the existence of a weak secondary population at the location of the WFS level (the bump of the solid curve), although its magnitude is quite small. Such a secondary population is also seen for the red line pair in the right panel, in the same relative position of the Gaussian wing, but its location is here much below the WFS level. This is again a consequence of the contamination of the red line ratio by thermodynamic effects. It is interesting that such a secondary population  is visible not only in observations with the high angular resolution of \textit{Hinode} \citep{Stenflo10, Stenflo11} but shows up also with our relatively low spatial resolution. 

In  Figure \ref{F6-1sigma-absV} we have plotted as the solid curves the standard deviation of the line ratio for a sequence of $\vert V/I\vert$ bins, as determined from the 1-$\sigma$ width of the fitted Gaussians, as a function of $\vert V/I \vert$. This result is compared with what one would expect from the measurement noise alone (dash-dotted lines), derived from the error bars of the determined $V/I$ values. We see that the agreement between the two curves is very good, there is no evidence that the total spread significantly exceeds the instrumental noise. This demonstrates that the spread around the regression lines that we see in the scatter-plot diagrams is practically exclusively instrumental, without evidence for any intrinsic solar spread. 

The tightness of the regression relation without any discernible solar spread beyond the measurement noise (which for ZIMPOL is very small) represents an important constraint that needs to be satisfied by any theoretical simulation of magneto-convection. It is important to remember that any realistic future numerical simulations must be able to reproduce not only the average line ratios, but also the astonishingly small solar spread of the line-ratio values. It may well turn out that the small spread is the constraint that is the most difficult one to satisfy, which would reveal serious deficiencies not uncovered by the other constraints.

\subsection{Dependence of the line ratio on flux density}\label{sec:fluxdep}
In Figure \ref{F7-rat-absV-wide} we plot the value of the line ratio as determined from the regression slope for a sequence of $\vert V/I \vert$ bins as a function of $\vert V/I \vert$, for the same disk-center data set as used for the previous figures. Since the line ratio varies very steeply for the smallest $V/I$ values (which represent the smallest flux densities), approaching the weak-field level in the limit of zero flux density, we need to use a logarithmic $\vert V/I \vert$ scale. For $\vert V/I \vert$ larger than a few promille the line ratio varies slowly and stays far below the weak-field level, indicating (in the case of the green line ratio) intrinsically strong (kG type), collapsed fields as the source of the $V/I$ signals. As before, the red line ratio is much more separated from the weak-field ratio as a consequence of the contamination from thermodynamic effects. This type of behavior (decrease of the line ratio with increasing of flux density) has also been noticed in observations with other instruments  \citep{Demidov-etal08,Stenflo10}.

\begin{figure}
\centering
\resizebox{0.9\hsize}{!}{\includegraphics{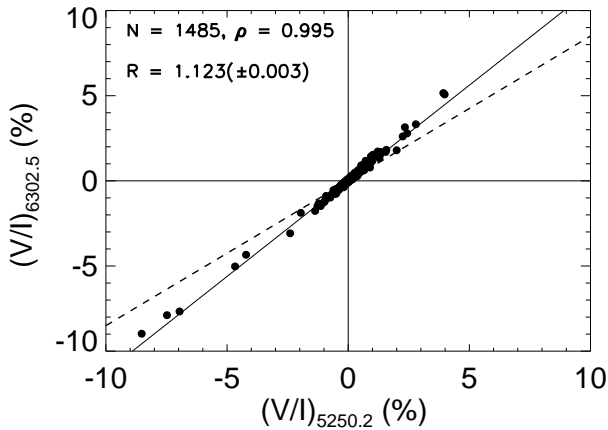}}
\caption{$V/I$ scatter plot of the red Fe\,{\sc i}~6302.5\,\AA\ line vs. the green Fe\,{\sc i}~5250.2\,\AA\ line, for an observational sample of 11 disk-center recordings in ``simultaneous'' mode, carried out at IRSOL on August 11, 2012. Both the larger scatter of the points around the regression line (solid) and its deviation from the WFS line (dashed) are due to the combined effects of differential Zeeman saturation and differential thermodynamic effects. 
}\label{F8-r2g2}
\end{figure}

\subsection{5250-6302 cross-calibration}\label{sec:crosscal}
Besides the derivation of line ratios for specially selected line pairs of the same atomic multiplet, a comparison between observations made in different individual spectral lines in general, and for  Fe\,{\sc i} 5250.2\,\AA\  and   Fe\,{\sc i} 6302.5\,\AA\ in particular, is also of importance.  There is for instance a need for mutual cross-calibrations between the various long-term  measurements in ground based observatories, where  the Fe\,{\sc i} 5250.2\,\AA\  line has been used for  many decades, and the recent space observations with \textit{Hinode}. Our recordings of the green and red line pairs in the simultaneous mode allow us to make scatter plots not only for each line pair, but also for various combinations of lines from different line pairs. As an example we show in Figure \ref{F8-r2g2}  the $V/I$ scatter plot of the 6302.5\,\AA\ line vs. the 5250.2\,\AA\ line, with the regression line fit (solid) and the corresponding weak-field relation (dashed). The spread of the points around the regression relation is now significantly larger than for the scatter plots of each separate line-ratio pair, and is therefore of solar origin. Its cause is the combined effect of differential Zeeman saturation and differential thermodynamic and line-formation effects in the two lines. This combined effect is also the cause for the deviation of the regression line (solid) from the WFS line (dashed).

\section{Observations with the STOP telescope}\label{sec:stop}

\begin{figure}
\centering
\resizebox{0.9\hsize}{!}{\includegraphics{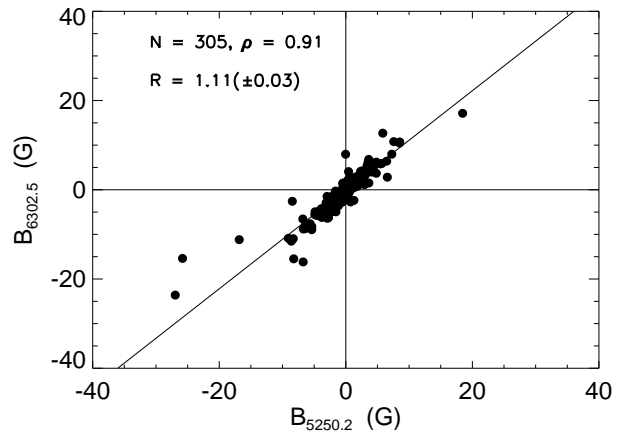}}
\caption{Scatter plot of the apparent flux density $B_{\rm 6302.5}$ for the 6302.5\,\AA\ line, derived from the measured Stokes $V$ assuming intrinsically weak fields, vs. the corresponding quantity $B_{\rm 5250.2}$ for the 5250.2\,\AA\ line. The data are taken from two consecutive full-disk magnetograms with about 100\,arcsec resolution in these two spectral lines, recorded with the STOP telescope of the Sayan Solar Observatory on March 26, 2012. 
}\label{F9-r2g2-stop}
\end{figure}
  
Some observations with the aim of exploring the relation between the apparent flux densities measured in the green and red line pairs that we have discussed here were also made with the  STOP telescope of the Sayan Solar Observatory. Since the green and red spectral domains could not be observed simultaneously (because the STOP spectrograph is not achromatic), it was decided to record the magnetograms  with low spatial resolution, about 100\,arcsec,  to minimize the influence of temporal changes between successive ``green'' and ``red'' magnetograms. The temporal separation between these two magnetograms as recorded on March 26, 2012, was about two hours. The resulting scatter plot for the apparent flux densities $B_{\rm 6302.5}$ vs. $B_{\rm 5250.2}$ is shown in Figure \ref{F9-r2g2-stop}.  Note that in contrast to the data from ZIMPOL the quantity that is plotted is not $V/I$ but the apparent flux density $B$, which is derived from the observed Stokes $V$ on the basis of the weak-field approximation. The regression-line slope therefore needs to be converted before it can be compared with the slope in the corresponding 6302.5-5250.2 scatter plot of Figure \ref{F8-r2g2}. 
 
If we divide the regression slope 1.123 of Figure \ref{F8-r2g2} by 0.85 to normalize it to the weak field slope (which is implicitly already done to obtain the flux densities $B$ in Figure \ref{F9-r2g2-stop}), we obtain 1.32, which differs significantly from the ratio 1.11 found in Figure \ref{F9-r2g2-stop}. Much of this discrepancy has to do with the center-to-limb variations (CLV) of the line ratios, which we will explore in  Sect.~\ref{S-CLV LR}. 

\begin{figure*}
\centering
\resizebox{0.9\hsize}{!}{\includegraphics{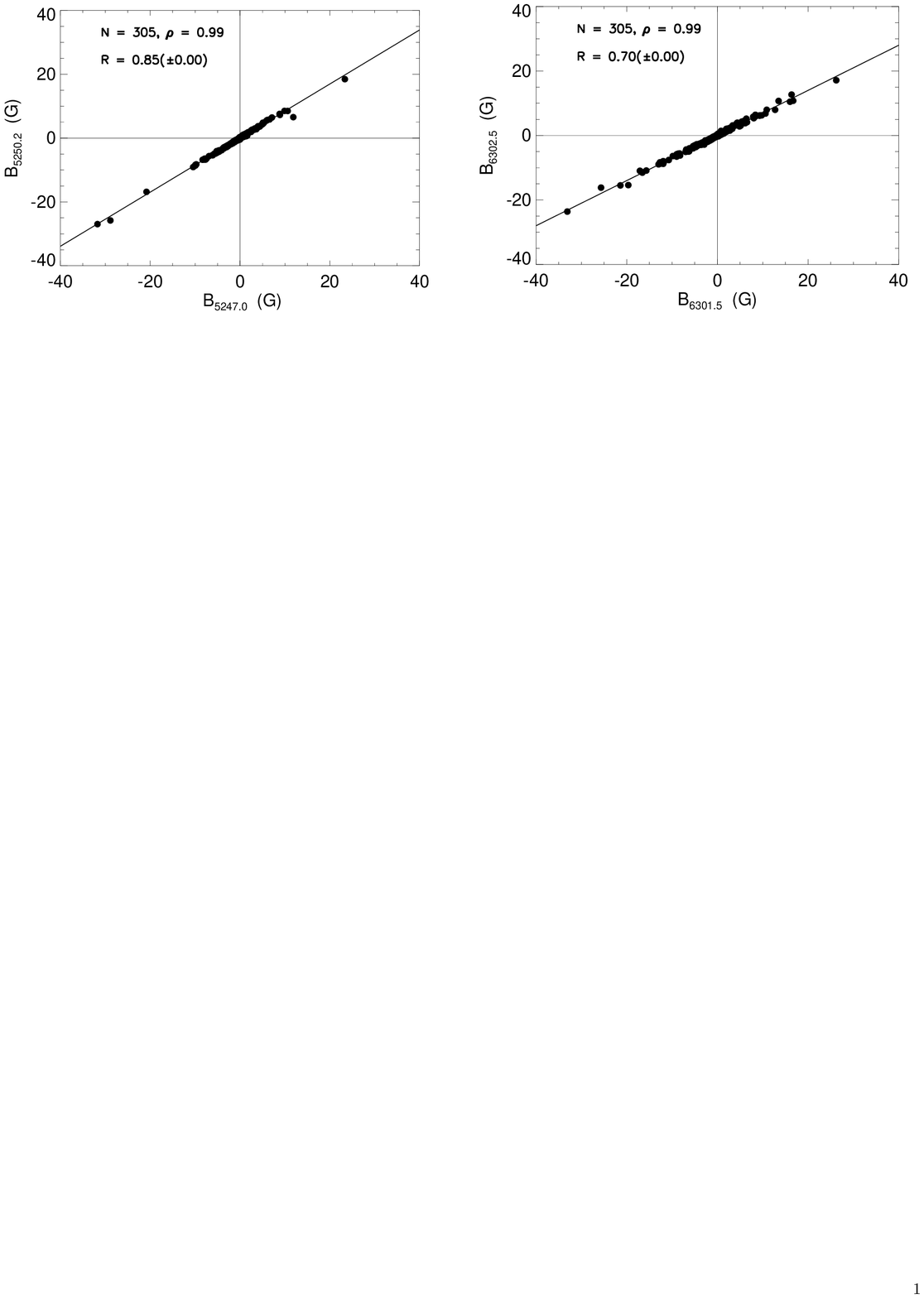}}
\caption{Scatter plots of the apparent flux densities measured simultaneously in the same full-disk magnetogram for the green line pair (left panel) and the red line pair (right panel). The recordings were  made at the STOP telescope of the Sayan Solar Observatory on March 26, 2012. Note the small scatter around the regression relations. 
}\label{F10-g2g1-r2r1-stop}
\end{figure*}

We notice in Figure \ref{F9-r2g2-stop} that in spite of the rather high correlation coefficient ($\rho = 0.91$), the scatter of the points is much larger than in Figure \ref{F8-r2g2}. There are various reasons for this. Firstly the two-hour temporal separation may cause some spatial mismatch (due to solar rotation) and evolutionary changes even with such low spatial resolution. A second cause is the mentioned center-to-limb variations of the line ratios. While the data in Figure  \ref{F9-r2g2-stop} represent full-disk observations, including limb regions, the data in Figure \ref{F8-r2g2} refer to the center of the solar disk. A third reason comes from the dependence of the line ratios on flux density (cf. Figure \ref{F7-rat-absV-wide}). 
  
This solar scatter indeed goes away when one instead makes scatter plots between line pairs recorded simultaneously in each full-disk magnetogram. This is shown in Figure \ref{F10-g2g1-r2r1-stop}  for the same STOP observations of March 26, 2012. The left panel shows the scatter plot for the green line pair, the right panel for the red line pair. Since in this case the line combination of each plot represents simultaneous observations, the issues of spatial and temporal mismatch disappear, which results in extremely tight regression relations, as we also found with ZIMPOL at IRSOL. 

\begin{figure*}
\centering
\resizebox{0.85\hsize}{!}{\includegraphics{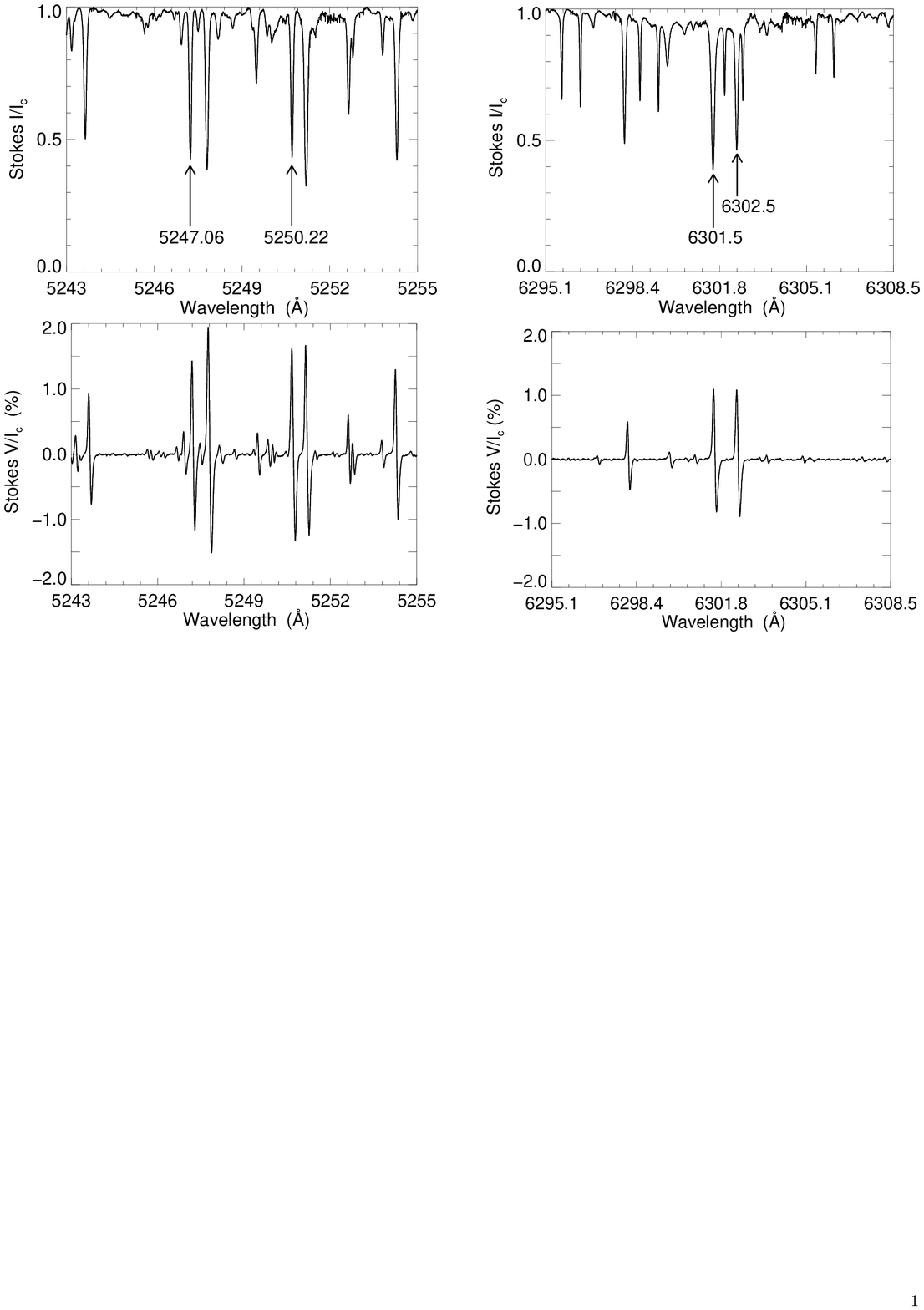}}
\caption{Example of Stokes $I/I_{c}$ (top panels) and $V/I_{c}$ (bottom panels) spectra for the green (left panels)  and red (right panels) regions as recorded by STOP at a spatial location  near disk center, where the 5250.2\,\AA\ magnetic flux density was about 20.5\,G.
}\label{F11-I-V-g-r-STOP}
\end{figure*}

\begin{figure*}
\centering
\sidecaption
\includegraphics[width=8.3cm]{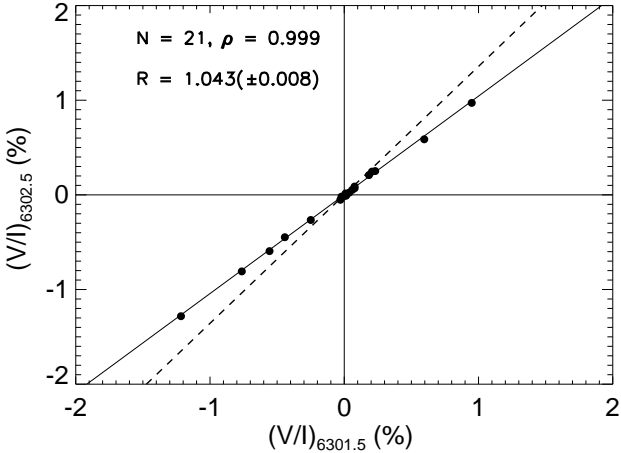}
\caption{$V/I$ scatter plot for the red line pair, based on STOP observations, but processed with the same algorithm that has been used for the ZIMPOL data sets.  
}\label{F12-r2-r1-as-IRSOL}
\end{figure*}

\begin{figure*}
\centering
\resizebox{0.9\hsize}{!}{\includegraphics{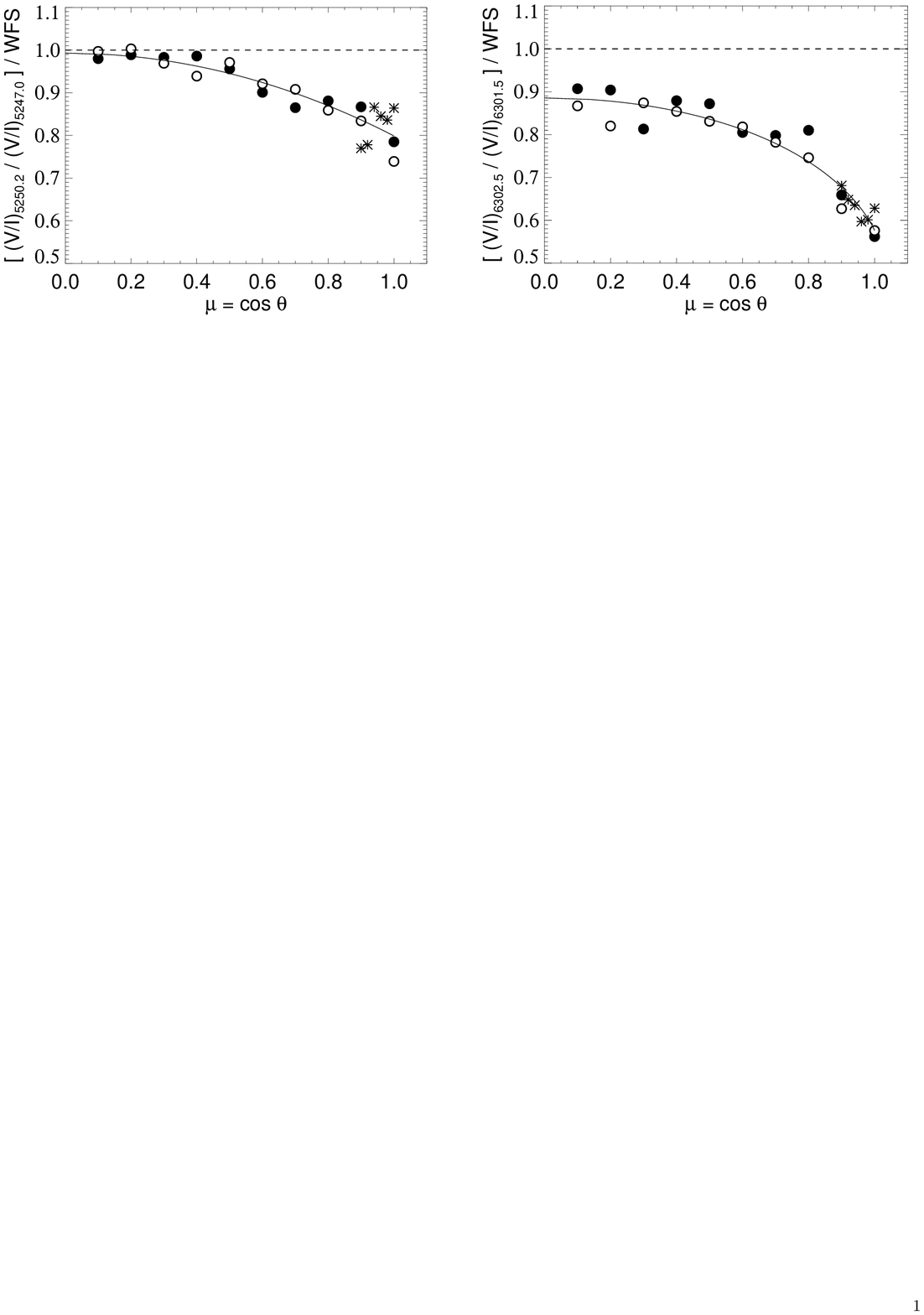}}
\caption{Center-to-limb variation (dependence on $\mu = \cos \theta$) of the $V/I$ line ratios. Left panel: green line ratio. Right panel: red line ratio. The filled circles represent observations in separate mode, the open circles and asterisks observations in simultaneous mode. The solid curves are analytical fit functions as described in the text. 
}\label{F13-mu-all-fit}
\end{figure*}

Let us compare the values for the regression slopes obtained with STOP and ZIMPOL. For a  translation of the ZIMPOL slopes, which are based on $V/I$ scatter plots, to the STOP slopes that are based on $B$ (flux density) scatter plots, the ZIMPOL slopes first need to be normalized with the corresponding WFS slope from the same scatter plot. For the green line pair we have very good agreement, since the STOP $R$ = 0.85 (cf. the left panel of Figure \ref{F10-g2g1-r2r1-stop}), while for ZIMPOL the normalized $R$ = 1.225/1.476 = 0.83 (cf. the left panel of Figure  \ref{F2-g2g1-r2r1-2filt}). For the red line pair the difference is a bit larger: $R$ = 0.70 for STOP and $R$ = 0.788/1.385 = 0.57 (right panel of Figure \ref{F2-g2g1-r2r1-2filt}) for ZIMPOL. A still larger discrepancy is found for the red-green 6302.5-5250.2 combination. In this case we have $R$ = 1.11 for STOP, $R$ = 1.123/0.85 = 1.32 (cf. Figure \ref{F8-r2g2}) for ZIMPOL. It should however be remembered that the STOP slopes are based on full-disk magnetograms, while the ZIMPOL values refer to disk center only. In view of the CLV of the line ratios and their flux density dependence, such differences between the STOP and ZIMPOL data sets are not surprising.  It is not clear why the difference is so much more pronounced for the red line ratio as compared with the green ratio, and if this difference in behavior is statistically significant. However, as will be seen in Fig.~\ref{F13-mu-all-fit} later, the CLV for the green ratio is shallower than that of the red ratio, and the green ratio is free from differential thermodynamic effects.
   
Figure \ref{F11-I-V-g-r-STOP} illustrates the high quality and low noise of the Stokes spectra recorded with STOP. The panels show the Stokes $I/I_{c}$  and $V/I_{c}$ spectra in the green and red ranges used for the present work. They represent one spatial point  near disk center  with magnetic flux density in the Fe\,{\sc i}~5250.2\,\AA\ line of about 20.5\,G.

 With the standard STOP reduction Stokes $V$ is normalized in terms of the continuum intensity $I_c$ in contrast to the ZIMPOL data, which uses the ``local'' Stokes $I$ instead.  Since the STOP spectra allows us to emulate the ZIMPOL reduction procedure, we have done this for 21 selected points in full-disk magnetograms to see if it makes any difference to the scatter plots and regressions. The 21 points, each representing Stokes spectra with 3585 pixels along the dispersion direction, have been selected such that the whole range of flux densities that are present in the right panel of Figure \ref{F10-g2g1-r2r1-stop} gets spanned. The result is shown in Figure \ref{F12-r2-r1-as-IRSOL} in terms of a $V/I$ scatter plot for the red line pair. 

The standard reduction procedure at STOP is to determine the magnetic flux density $B$ by the center-of-gravity method. After normalization of the regression slope in Figure \ref{F12-r2-r1-as-IRSOL} in terms of the weak field slope (which equals 1.5), we obtain the regression slope in terms of flux densities as 1.043/1.50 = 0.70, which is exactly the same value as found in Figure \ref{F10-g2g1-r2r1-stop}  (right panel). This demonstrates that the difference in reduction technique has no significant influence on the slope values. For the red line pair 1\,\%\ of circular polarization corresponds to approximately 20 G.

\section{Center-to-limb variations of line ratios and calibration of the 6302/6301 line ratio}\label{S-CLV LR}
Since the Stokes $V$ line ratios are known to vary significantly with center-to-limb distance \citep[cf.][]{HowardStenflo72,Stenflo-etal87}, we did a number of observing runs with ZIMPOL at IRSOL for a sequence of center-to-limb positions, spanning the $\mu$ (cosine of the heliocentric angle) range from 1.0 (disk center) to 0.1 (5\,arcsec inside the solar limb). We did such observing runs for the green and red line pairs both in separate mode and in simultaneous mode (when the green and red line pairs are recorded side by side on the same CCD sensor). 
     
Figure \ref{F13-mu-all-fit} shows the results of several  independent center-to-limb observing runs, for the separate mode as filled circles, for the simultaneous mode as open circles. Since the near disk-center range $\mu=0.9$-1.0 is rather important for the interpretation of the green line ratio in terms of flux tube models \citep{Solanki1998}, we have made a special run (on August 28, 2012) to have detailed coverage of this $\mu$ range, in steps of $\Delta \mu = 0.02$, for the mode with simultaneous recording of the two line pairs on the same CCD. The results of this run are plotted as the asterisks in Figure \ref{F13-mu-all-fit}. They demonstrate that the $\mu$ dependence in this narrow $\mu$ range is relatively smooth and not abrupt. Much of the scatter of the points is caused by the $V/I$ amplitude dependence of the line ratio (cf. Figure \ref{F7-rat-absV-wide}). Since the different sampled spatial positions have different flux density distributions, the average line ratio is formed with different flux-density weightings, which leads to different average results. 

While the relative shape of the $\mu$ dependence is similar for the green and red line ratios, the red line-ratio curve is displaced to values further below the WFS (weak-field slope) reference level (dashed line). The green line ratio approaches the weak-field case when going to the limb, while the red line ratio remains much below the WFS also at the extreme limb, again a consequence of its ``contamination'' by differential thermodynamic effects. 

\begin{figure}
\centering
\resizebox{\hsize}{!}{\includegraphics{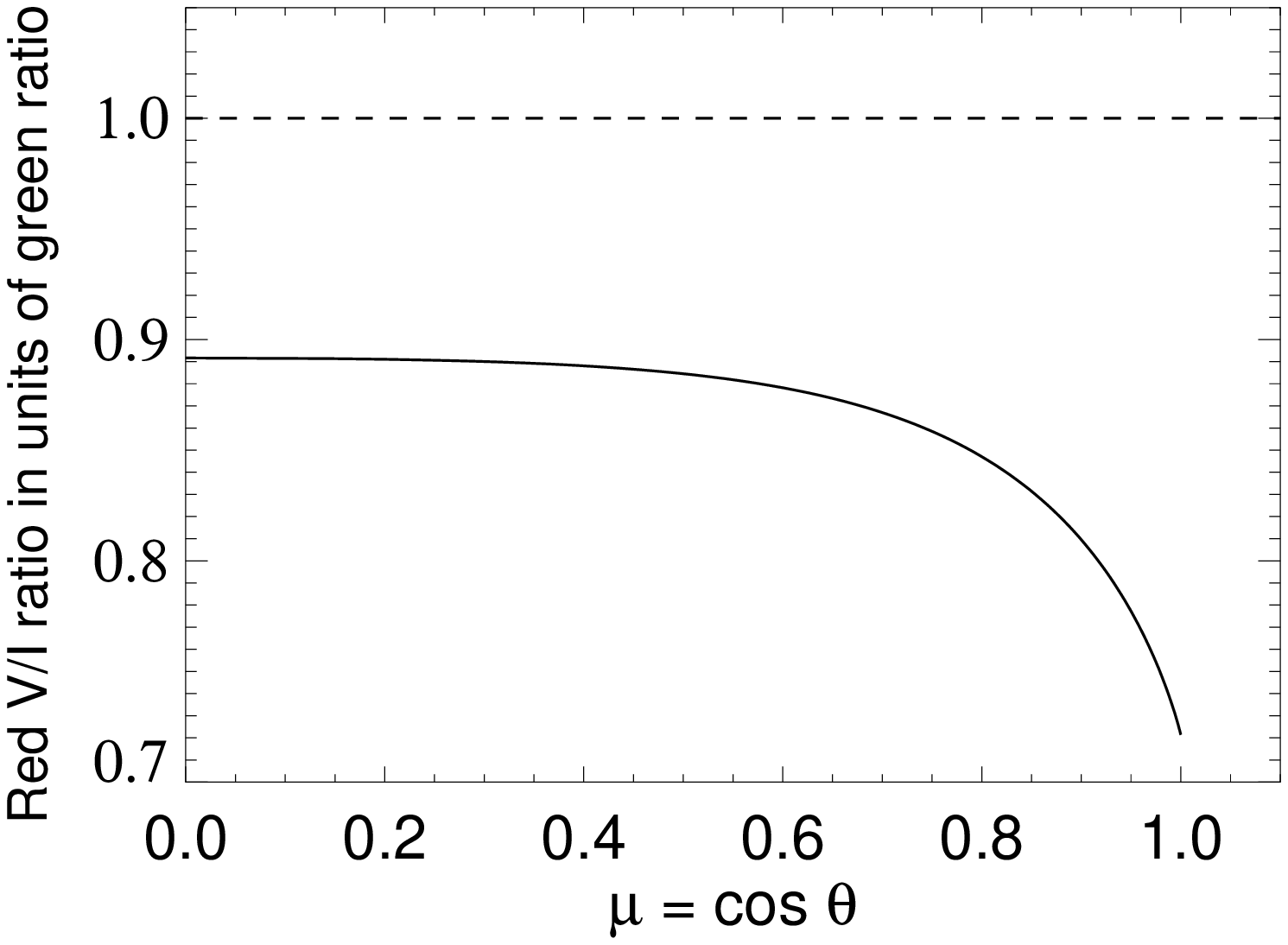}}
\caption{Analytic representation of the red $V/I$ line ratio in units of the green ratio as a function of $\mu$, obtained by forming the ratio between the red and green fit curves of  Figure \ref{F13-mu-all-fit}.  
}\label{F14-red-gr-fit}
\end{figure}

\begin{figure*}
\centering
\resizebox{0.9\hsize}{!}{\includegraphics{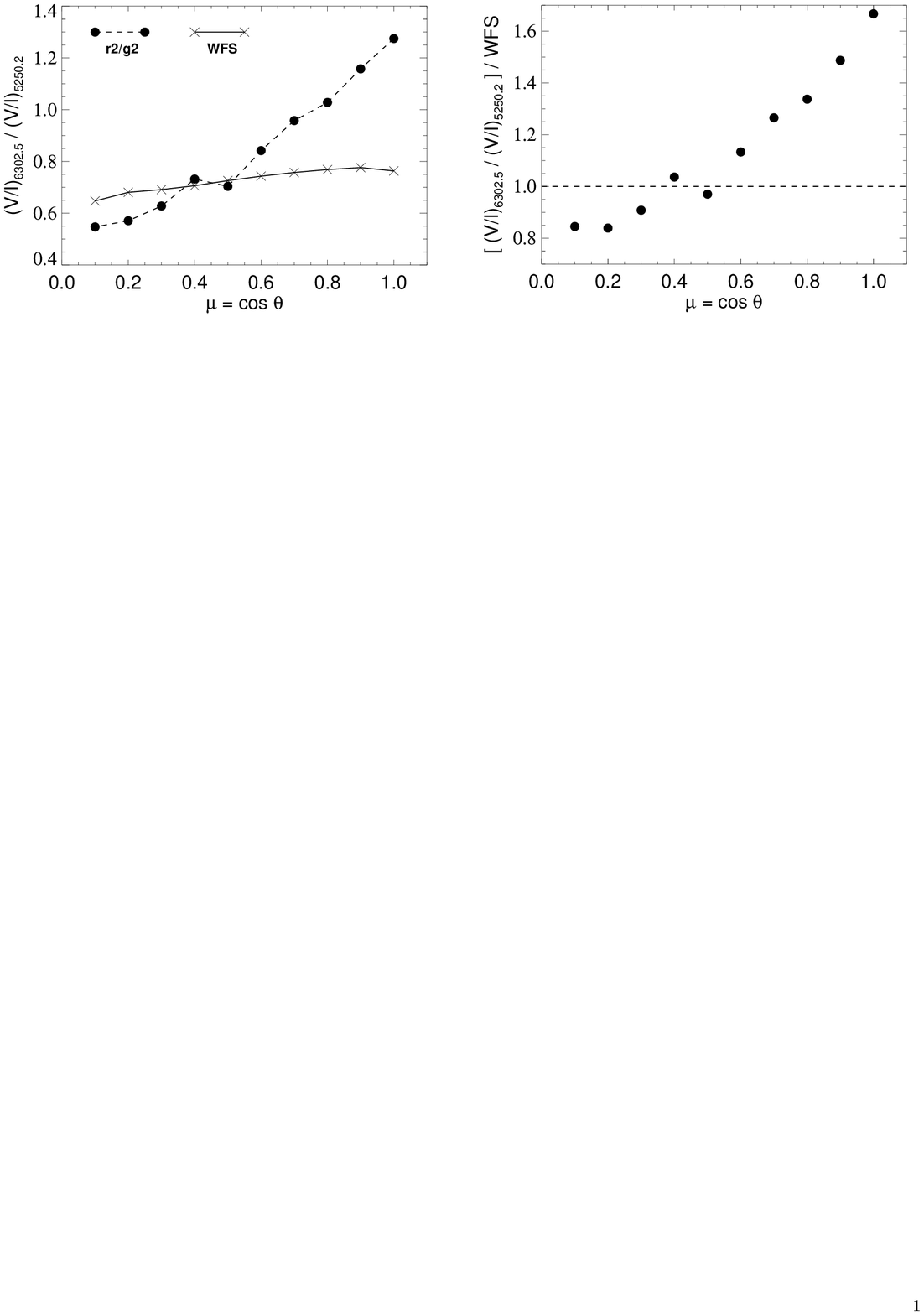}}
\caption{$(V/I)_{6302.5}/(V/I)_{5250.2}$ ratio for the red-green line combination, based on ZIMPOL observations in simultaneous mode, together with the weak-field slope WFS in the left panel, as functions of center-to-limb position given by $\mu$. In the right panel the line-ratio values have been normalized in terms of the WFS. 
}\label{F15-r2g2-mu}
\end{figure*}

The solid curves in Figure \ref{F13-mu-all-fit} are analytical functions used to fit the data. Through trial and error we found that the line ratio $R$, when normalized to the WFS level $R_{\rm WFS}$ as in Figure \ref{F13-mu-all-fit}, can be well represented by the function $R/R_{\rm WFS}=a_0 + a_1x\,[\,1-\exp(-a_2x)\,]$\,, where $x=r/r_\odot =\sqrt{1-\mu^2}$ is the radius vector in units of the radius of the solar disk. $a_{0,1,2}$ are the three free parameters that are determined by iterative least-squares fitting, with all 26 points in each of the two panels of Figure \ref{F13-mu-all-fit} being given equal weight. For the green line-ratio curve $a_{0,1,2}=0.799$, 2.98, and 0.067, respectively, while for the red line-ratio curve these parameter values are 0.576, 0.329, and 2.77.  
  
By forming the ratio between the red and green fit curves represented in Figure \ref{F13-mu-all-fit}, we obtain the red $V/I$ line ratio expressed in units of the corresponding green line ratio. This curve tells how the contaminated red line ratio can be translated into the corresponding uncontaminated green ratio, i.e., it represents a calibration of the red ratio in terms of the green ratio. This ``calibration curve'' is shown in Figure \ref{F14-red-gr-fit} as a function of $\mu$.  If the red line ratio were uncontaminated, we would expect this curve to be approximately flat and very close to unity. In contrast the curve lies systematically well below unity and decreases steeply to much smaller values as we approach disk center. Through ``renormalization'' of the red line ratio with the help of this calibration curve it is possible to interpret the red line ratio in terms of intrinsic field strengths, which is not possible otherwise. 

The need for such a renormalization became evident in the analysis of \textit{Hinode} SOT/SP observations of the quiet-sun disk center \citep{Stenflo10}, as revealed by the huge discrepancy between the theoretically predicted $V_{\rm 6302.5}/V_{\rm 6301.5}$ ratio of 2.1 in the weak-field limit and the observed line ratio of 1.55 for the secondary weak-field population. This led to a mismatch factor of $1.55/2.1=0.74$, which represents the scale error of the red line ratio at disk center. It had to be renormalized away to allow extraction of  intrinsic field strengths from the observed red line ratio \citep[for details, see Sect. 6 of][]{Stenflo10}. At that time this renormalization procedure may have seemed a bit ad hoc, but the present work validates it: The renormalization factor of 0.74 is in nearly perfect agreement with the disk-center value of the calibration curve in Figure \ref{F14-red-gr-fit}. 

Let us recall that all the regression line slopes that enter into the line ratios used for Figs.~\ref{F13-mu-all-fit} and \ref{F14-red-gr-fit} have been determined with the assumption of a linear regression, thus disregarding the $V/I$ dependence that was shown in Fig.~\ref{F7-rat-absV-wide}. The scatter of the points around the analytical fit functions in Fig.~\ref{F13-mu-all-fit} cannot be explained in terms of measurement noise alone but most likely has its main origin in the disregard of the nonlinear aspect of the regression relation. In this sense the calibration curve in Fig.~\ref{F14-red-gr-fit} represents an approximate relation that ignores higher-order effects and may therefore be improved on in the future when the nonlinearities are properly included. 

If the linear regression assumption would continue to be used in future determinations of the line-ratio slopes, one might find an apparent solar-cycle variation of the calibration curve that could be an artefact due to the different weightings of the larger and smaller flux densities during different phases of the cycle. The question whether there could also be an intrinsic solar-cycle variation of the calibration curve is certainly of considerable interest, but this question is only answerable if the intrinsic dependence may be separated from spurious effects. Such a synoptic program would only have a chance of success if one can include full nonlinear regression modeling with large statistical samples.

Our simultaneous-mode observations with ZIMPOL of the two line ratios allows us to cross-compare and form ratios between any combination of the four lines. In Figure \ref{F15-r2g2-mu} we show the center-to-limb variation of the red/green ratio $(V/I)_{6302.5}/(V/I)_{5250.2}$ between the 6302.5 and 5250.2\,\AA\ lines, which indicates how a magnetogram in one of these two lines can be translated into a magnetogram in the other line. We find that this line ratio decreases by approximately a factor of two when going from disk center to the limb. The strong center-to-limb variation may be the main reason for the relatively large scatter of the points in Figure \ref{F9-r2g2-stop} for this line combination for the full-disk observations with the STOP telescope. We also note in Figure \ref{F15-r2g2-mu} that the CLV of the line ratio crosses the level that corresponds to the weak field case (although the local change of slope near this crossing is not statistically significant). This behavior cannot be interpreted directly, since it is subject to a mixture of various effects.

\section{Discussion and Conclusions}\label{S-Discussion}
The exploration of quiet-sun magnetic fields is particularly challenging because the  fractal-like magnetic structuring continues at scales far smaller than the resolved ones. The apparent flux densities that are measured have little direct relation with the intrinsic field strengths. On top of this the polarization signals found on the quiet Sun are tiny, which places great demands on the polarimetric sensitivity of the instruments used. 

Under these circumstances one needs to develop and apply robust diagnostic techniques that can provide information on the ``true'' underlying nature of the magnetic field, ``true'' in the sense that the conclusions are independent of the angular resolution of the telescope that is available at the time. Such a robust and resolution-independent method was provided by the 5250/5247 magnetic line ratio, with which it was possible to separate the magnetic-field effects from the thermodynamic and line-formation effects, allowing the discovery forty years ago that most of the magnetic flux seen in solar magnetograms has its origin in strong (of order kG) subresolution intermittent field-line bundles of collapsed flux \citep{Stenflo73}. With the significant advance in telescope resolution, in particular with the \textit{Hinode} space observatory \citep{Suematsu-etal08}, the line-ratio technique could again be applied to reveal the existence of a secondary, weak and uncollapsed flux population \citep{Stenflo10,Stenflo11}. However, a serious obstacle to the application of the line-ratio technique for \textit{Hinode} data is that the 5247-5250 line pair is not available but only the Fe\,{\sc i}~6301.5 and 6302.5\,\AA\ line pair, although it has been well known that the Stokes $V$ line ratio formed with this line combination is severely contaminated by thermodynamic and line-formation effects. The 6302.5/6301.5 line ratio could only be used for quantitative analysis of  the properties of the collapsed and uncollapsed flux populations after a special ``renormalization'' procedure had been applied to ``decontaminate'' it  \citep{Stenflo10}. The main aim of the present paper has been to explore the validity of this renormalization and in the same context allow the 6302.5/6301.5 line ratio to be calibrated in terms of the uncontaminated 5250/5247 line ratio. 

For this purpose we modified the spectrograph used by the ZIMPOL-3 imaging polarimeter at IRSOL such that both line pairs could be recorded simultaneously, side by side, on the same CCD sensor of the ZIMPOL camera system. This gave us the option of recording the two line pairs either in simultaneous or in separate mode. Observing runs in both modes were done to give us detailed determinations of these line ratios as functions of both flux density and center-to-limb distance. In addition complementary observations were carried out with the STOP telescope of the Sayan Solar Observatory in the form of full disk solar magnetograms in these line pairs. 

Analysis of line-ratio histograms recorded with ZIMPOL reveals the existence of a small secondary flux population that according to the 5250/5247 line ratio represents intrinsically weak fields. This secondary flux population, which showed up more prominently in the \textit{Hinode} line-ratio analysis, is thus discernible even with the much lower angular resolution at IRSOL. 

Similar to the findings with \textit{Hinode} \citep{Stenflo10}, both Stokes $V$ line ratios are found to vary significantly with apparent flux density. For most flux densities the line ratio stays well below the ratio that would be expected for intrinsically weak ($\la 500$\,G) fields, which in the case of the green (5250/5247) line ratio constitutes a signature for kG-type fields. In the limit of small flux densities the line ratio steeply approaches the weak-field level for both line pairs. The red (6302.5/6301.5) line ratio however deviates much more from the weak-field case as a result of its contamination by thermodynamic effects. 

As we move from disk center to the solar limb, the green line ratio approaches and reaches the weak-field level. The shape of the $\mu$ dependence is similar for the red line ratio, but since it is systematically displaced to lower values, it never reaches the weak-field level at the limb but remains well below. 

By forming the ratio between the center-to-limb curves for the two line ratios and expressing it in terms of an analytical fit function, we obtain the red line ratio in units of the green line ratio as a function of $\mu$. With this ``calibration curve'' in Figure \ref{F14-red-gr-fit} the contaminated 6302.5/6301.5 Stokes $V$ line ratios can be converted into the corresponding uncontaminated 5250/5247 line ratios for any center-to-limb position. It is this conversion that constitutes the ``renormalization'' that allows the 6302.5/6301.5 ratio to be quantitatively interpreted in terms of intrinsic field strengths. 

The \textit{Hinode} disk-center analysis of quiet-sun magnetic fields showed that the 6302.5/6301.5 Stokes $V$ ratio was too small by a factor of 0.74 to give a consistent interpretation in terms of intrinsic field strengths and in particular to allow the secondary flux population to be interpreted in terms of uncollapsed, intrinsically weak fields \citep{Stenflo10}. The \textit{Hinode} line-ratio data were therefore renormalized by this factor. According to the calibration curve in Figure \ref{F14-red-gr-fit} this factor is about 0.72 at disk center, in almost perfect agreement with the corresponding factor of 0.74 from \textit{Hinode} when considering the measurement uncertainties. This result validates the renormalization procedure used in the \textit{Hinode} analysis.

\begin{acknowledgements}
The results obtained in the present work have been obtained partly through the financial support of the Scientific \&\ Technological Cooperation Programme Switzerland-Russia, which allowed a three-month visit of MLD to IRSOL in the summer of 2012. The research at IRSOL is financially supported by Canton Ticino, the Swiss Confederation (SER), the city of Locarno, the local municipalities, and SNF grant 200020-127329. RR acknowledges financial support from the Carlo e Albina Cavargna Foundation.
\end{acknowledgements}

\end{document}